\documentclass[10pt,twocolumn]{article}

\usepackage[utf8]{inputenc}
\usepackage{cvpr}
\usepackage{times}
\usepackage{epsfig}
\usepackage{graphicx}
\usepackage{amsmath}
\usepackage{amssymb}
\usepackage{color}
\usepackage{graphbox}
\usepackage[english]{babel}
\usepackage{caption}
\usepackage{subcaption}
\definecolor{teal}{rgb}{0,0.5,0.5}
\definecolor{darkgreen}{rgb}{0,0.5,0}

\usepackage[pagebackref=true,breaklinks=true,colorlinks,bookmarks=false]{hyperref}

\cvprfinalcopy

\begin{document}

\def\L{l}
\def\C{c}
\def\M{m}

\def\S{S}
\def\Lset{L}
\def\Cset{C}
\def\Mset{M}
 
\title{A Calibration Scheme for Non-Line-of-Sight Imaging Setups}

\author{Jonathan Klein\\
{\small University of Bonn}\\
{\tt\small kleinj@cs.uni-bonn.de}
\and
Martin Laurenzis\\
{\small French-German Research Institute of Saint-Louis}\\
{\tt\small martin.laurenzis@isl.eu}
\and
Matthias B. Hullin\\
{\small University of Bonn}\\
{\tt\small hullin@cs.uni-bonn.de}
\and
Julian Iseringhausen\\
{\small University of Bonn}\\
{\tt\small iseringhausen@cs.uni-bonn.de}
}

\twocolumn[{
\renewcommand\twocolumn[1][]{#1}%
\maketitle

\vspace{-1.5em}
\noindent\begin{minipage}{\textwidth}
\center
\begin{minipage}{0.25\textwidth}
  \centering
  \includegraphics[width=\linewidth]{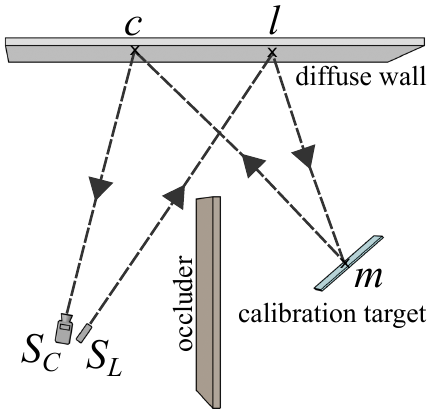}
  \captionof{subfigure}{Scene setup}
  \label{fig:teaser_overview}
\end{minipage}
\begin{minipage}{0.34\textwidth}
  \centering 
  \includegraphics[width=\linewidth]{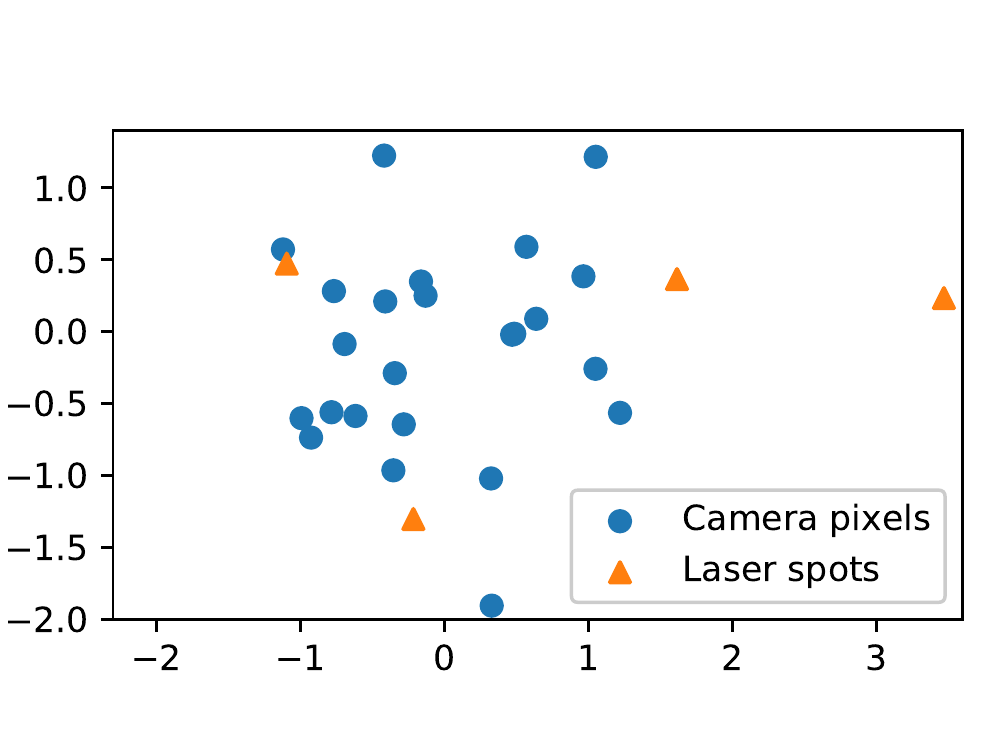}
  \captionof{subfigure}{Initialization}
  \label{fig:teaser_init}
\end{minipage}
\begin{minipage}{0.34\textwidth}
  \centering 
  \includegraphics[width=\linewidth]{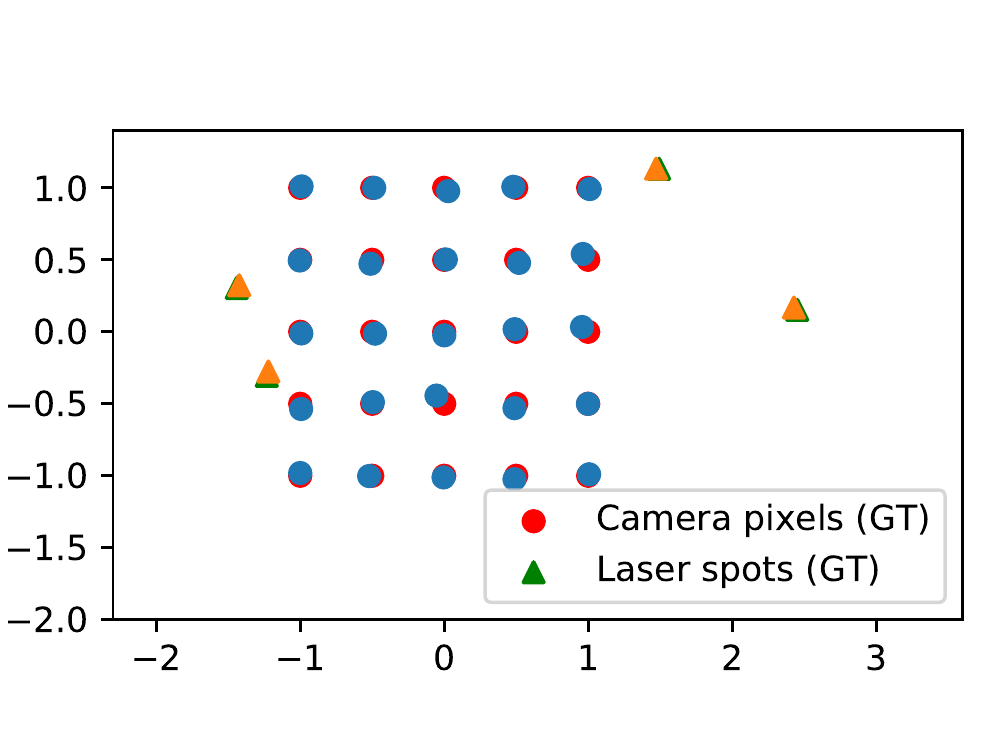}
  \captionof{subfigure}{Calibration result}
  \label{fig:teaser_solution}
\end{minipage}
\captionof{figure}{
(a)
We propose a novel method for the geometric calibration of three-bounce non-line-of-sight setups using transient imaging hardware.
Light travels from a laser $\S_L$ to a laser spot $\L$ located on the diffuse reflector wall.
From there, it is reflected towards a calibration target $\M$ and back to a projected camera pixel $\C$, finally reaching the camera $\S_C$.
We calibrate the setup using multiple images of a specular, planar mirror in different positions and orientations, analog to the procedure in classical 2D camera calibration.
Instead of relying on known features on the calibration target, we use the time of flight of the full path from laser to camera to solve for the individual laser spot positions $\L$ and projected camera pixels $\C$.
(b) The optimization problem is non-convex and requires a coarse initialization.
(c) Even in the presence of time-of-flight noise, our method reconstructs the acquisition setup geometry up to a very high precision.
The ground truth values (shown in red and green) are barely visible under the reconstruction.
}
\label{fig:teaser}
\vspace{1.5em}
\end{minipage}
}]

\begin{abstract}
The recent years have given rise to a large number of techniques for ``looking around corners'', i.e., for reconstructing occluded objects from time-resolved measurements of indirect light reflections off a wall. While the direct view of cameras is routinely calibrated in computer vision applications, the calibration of non-line-of-sight setups has so far relied on manual measurement of the most important dimensions (device positions, wall position and orientation, etc.). In this paper, we propose a semi-automatic method for calibrating such systems that relies on mirrors as known targets. A roughly determined initialization is refined in order to optimize a spatio-temporal consistency.
Our system is general enough to be applicable to a variety of sensing scenarios ranging from single sources/detectors via scanning arrangements to large-scale arrays. It is robust towards bad initialization and the achieved accuracy is proportional to the depth resolution of the camera system.
We demonstrate this capability with a real-world setup and despite a large number of dead pixels and very low temporal resolution achieve a result that outperforms a manual calibration.

\end{abstract}

\section{Introduction}
The ability to ``see'' beyond the direct line of sight forms an intriguing problem to be solved that is of great interest for many applications. These range from emergency situations, where situational awareness about dangers and victims is key, 
to scientific scenarios, where microscopes supporting such techniques reveal hidden structures.
The recent years have produced a number of techniques that sense objects located ``around a corner'' by recording time-resolved optical impulse responses, where light that bounces off a directly visible wall enters the occluded part of the scene and thus gathers information about hidden objects; see Figure~\ref{fig:teaser}\textcolor[rgb]{1,0,0}{a} for a schematic illustration. 
The available operation modes~\cite{Velten:2012:Recovering,Kirmani:2009,Heide:2019:OcclusionNLOS,OToole:2018,gariepy2015single,iseringhausenNlos2019} support not only object detection and tracking of components of the occluded scene but extend to the full reconstruction of 3D shape and texture.
In general it is assumed that the entire geometry of the setup is known and only the hidden object is to be reconstructed. This implies that the capture must be preceded by a manual calibration: Positions and distances of devices and objects have to be measured with high accuracy, a task which is tedious and often results in imprecise results. 

Here, we propose what we believe is the first automatic system for calibrating the geometry of non-line-of-sight sensing setups. Our scheme does not require any additional hardware other than a common, planar mirror which serves as the calibration target. As in traditional camera calibration, the target is recorded in different positions and orientations. Since it does not rely on the target being textured, and since only a temporal onset (rather than the full time-of-flight histogram) is used, our calibration scheme can be employed for all types of ultrafast sensors, including single-pixel sensing scenarios~\cite{musarra2019singlepixel}, randomly scattered measurement locations~\cite{Buttafava:2015} as well as low-resolution imagers and even correlation time-of-flight sensors~\cite{Heide:2014}. Additionally, task-specific constraints (e.g., pixel positions restricted to a scan line) are easily integrated in the method.

Our calibration scheme requires an initialization to warm-start the non-linear optimization problem. 
In contrast to a laborious measurement, however, we rely only on a rough estimate of the setup's geometry:
As long as the initial solution is reasonably close to the true geometry, the method is robust to large amounts of statistical error like shot noise.

On a experimental measurement setup, we demonstrate that our scheme not only recovers relevant parameters to high accuracy, but that it also improves the outcome of NLoS reconstructions obtained using data from the setup.

\section{Related Work}
The last decade has given rise to a comprehensive body of work on non-line-of-sight sensing, i.e., the estimation of targets hidden from direct view by means of light undergoing indirect diffuse scattering off directly visible proxy objects. While various lines of research are exploring the use of steady-state measurements in order to extend the direct line of sight~\cite{katz2012looking,klein2016tracking,bouman2017turning,thrampoulidis2018exploiting,seidel2019,chen_2019_nlos}, the majority of works remains focused on the use of time-resolved measurements (transient images). 

A survey by Jarabo et al.~provides a good overview of transient imaging as per 2017~\cite{jarabo2017recent}. Seminal works include the recovery of low-parameter geometry and reflectance models from transient measurements~\cite{Kirmani:2009,Naik:2011} as well as the first reconstruction of distinct shapes as demonstrated by Velten et al.~\cite{Velten:2012:Recovering}. 
Since then, significant effort has been devoted to unlock novel sensor technologies and interferometric setups for transient imaging~\cite{Heide:2013:LBT,gariepy2015single,Gkioulekas:2015:MLT:2809654.2766928} while simultanously improving the performance of the de-facto standard reconstruction technique, ellipsoidal error backprojection~\cite{Velten:2012:Recovering,laurenzis2014feature,ArellanoOpEx2017}. 
Recent additions to the non-line-of-sight reconstruction problem include the introduction of the confocal capture setting~\cite{OToole:2018} as well as attempts to cast the problem into paradigms borrowed from wave optics and seismic tomography~\cite{liu2019non,Lindell:2019:Wave}. 
While most of these works rely on volumetric representations for the hidden target, other researchers have explored alternative, surface-driven representations as well \cite{pediredla2017reconstructing,iseringhausenNlos2019,tsai2019beyond}. 
These models typically lead to improved consistency of the solution with respect to a physically-based forward simulation of light transport, and they also naturally express effects like surface reflectance (BRDFs) or self-occlusion. 
Equipping volumetric representations with such surface-based characteristics to ``guide'' the reconstruction is possible, 
but comes at greatly increased implementation effort and computational cost~\cite{Heide:2014,Heide:2019:OcclusionNLOS}.\\
\newline
All these approaches share the need for a highly accurate calibration of the setup. Not only do the intrinsic and extrinsic parameters of capture devices have to be known, but so do the shape and orientation of proxy surface(s) that are to serve as ``mirrors'' into the occluded part of the scene. 
The state of the art still relies on the manual measurement of positions and relative distances. In direct imaging on the other hand, much more advanced and accurate, and at least semi-automatic, techniques are routinely used. 
For instance, for computational photography and computer vision purposes, intrinsic camera parameters are estimated by enforcing consistency between model prediction and observation for known reference targets like planar checkerboard patterns~\cite{zhang1999}. 
With this work we introduce a similar methodology for non-line-of-sight sensing setups, aiming to reduce the gap between direct and indirect imaging techniques with respect to accuracy, 
usability and robustness of setup calibration that has existed to this day.
 
\section{Method}

A non-line-of-sight (NLoS) setup can be viewed as a high-dimensional function that produces measurements from parameters such as the setup geometry, the hidden object, reflective properties of various components, a background signal, the sensor model of the camera, and others.  We distinguish radiometric parameters (that govern the \emph{amount} of light being transported) and spatio-temporal parameters (that govern the time of flight).
A first abstraction step drops camera and laser peculiarities and describes measurements as transient histograms, i.e., the time-resolved (on a pico- to nanosecond scale) intensity of light arriving at each sensor pixel. 
Commonly all participating reflectance functions (BRDF) are assumed to be Lambertian (with notable exceptions such as NLoS BRDF reconstruction \cite{Naik:2011} or retro-reflective objects \cite{OToole:2018}), and scenes are set up to minimize reflections from the background. 
This leaves only the setup and hidden object geometry, i.e., the spatio-temporal parameters of the system. As the image formation model of transient light transport is well understood (for a list of simulation approaches see Jarabo et al.~\cite{jarabo2017recent}) an analysis-by-synthesis approach can be employed to reconstruct the hidden geometry~\cite{iseringhausenNlos2019, klein2016tracking} but it relies on accurate knowledge of the setup.

A setup calibration can in principle be attempted in a similar fashion: Given a known hidden object the setup is inferred from measured transient data. 
However providing ground truth information for hidden objects is a challenging task: The potentially complex object geometry must be either measured or manufactured precisely and the object must be placed at one or multiple known positions. 
Thus it is unlikely that such an approach will result in fewer measurements than a manual setup calibration. Although in an even more general problem formulation a joint optimization of hidden object and setup geometry could be attempted, this seems not very promising given that reconstructing just the hidden object is already a challenging problem on its own.

Instead of using a traditional NLoS setup we therefore propose to replace the hidden object by a simple planar mirror (as available as common household object). As we will show in the following this makes the image formation model significantly simpler. This leads to an easier-to-solve optimization problem that has far weaker requirements on its initialization due to its implicit constraints. 
Our approach jointly optimizes for setup geometry and mirror placement, which allows for a setup calibration with little manual measurements that can be performed with reduced accuracy to acquire only a rough estimate for initialization.

The mirrors can be placed in the visible and hidden part of the scene. Thus access to the hidden part is not strictly required, however it can lead to more robust calibration, if it is accessible.

\subsection{Image Formation Model using Mirrors}
Figure~\ref{fig:teaser}\textcolor[rgb]{1,0,0}{a} gives a schematic illustration of a NLoS calibration setup: A sensor/laser light source setup on the left hand side which is separated from the mirror calibration target by an occluder. 
We denote the physical position of the camera and the laser with $\S_C$ and $\S_L$ respectively. 
As they are usually close to each other we define the shorthand notation $\S = \left\{ \S_C, \S_L\right\} $. In the classic three-bounce setup the signal is reflected from a planar wall. 
We denote the projected camera pixels on this wall with $\C \in \Cset$ and the (potentially multiple) laser spots with $\L \in \Lset$.  The mirrors that replace the hidden object in our setup are denoted with $\M \in \Mset$. 

$\Cset$ and $\Lset$ describe the setup as a very general combination of multiple camera pixels and laser positions. Whether the pixels lie on a fixed grid (as for 2D image sensors), a single line (as for streak cameras) or are placed arbitrarily on the wall (as for scans with single-pixel detectors) matters only insomuch as that some cases allow for specialized parameterizations that can improve calibrations (see \ref{sec:parameterization}). Due to the Helmholtz reciprocity the role of $\Lset$ and $\Cset$ are furthermore always interchangeable in the following discussion. 
Most common NLoS setups assume that all $\L \in \Lset$ and $\C \in \Cset$ lie on the same plane, which is the case for a planar wall. 
However, our method also works with general 3D points (and thus rough or curved walls), which allows us to cover a wide variety of NLoS setups that can be calibrated with our approach.

As the mirrors are convex, interreflections within the object, which usually blur the temporal response of the hidden object, are excluded implicitly.
In contrast, the specular reflection on the mirror allows for only a unique optical path $\L  \rightarrow \M \rightarrow \C$, connecting laser spot, mirror and projected pixel. 
Compared to classical transient rendering this means that no integration over the surface of the object is required, which allows for fast and noise-free computation. Our transient histograms only contain a single, sharp peak (when camera sensor model and laser characteristics are ignored). 
Given an intrinsic camera and laser calibration we assume that those peaks can be retrieved in a hardware-specific pre-calibration step that also deals with effects such as background radiation or higher-order bounces.

A complete measurement consists of a series of paths $P_{i,j,k} = \S_L \rightarrow \L_i  \rightarrow \M_j \rightarrow \C_k \rightarrow \S_C$ (we omit indices in unambigious cases). The image formation model assumes that only a single laser and a single mirror is used at a time.

Each path is characterized by a time-of-flight and an intensity. The intensity depends on the BRDF of the wall and its normal vector, while the time-of-flight is independent from both. For our calibration we only rely on the time-of-flight. We thus neither need to assume nor to estimate any BRDFs or wall normals (however the wall normal can still be retrieved for planar walls and estimated for rough or curved walls from the individual camera and laser points).

\begin{figure}
\begin{centering}
\includegraphics[width=0.8\columnwidth]{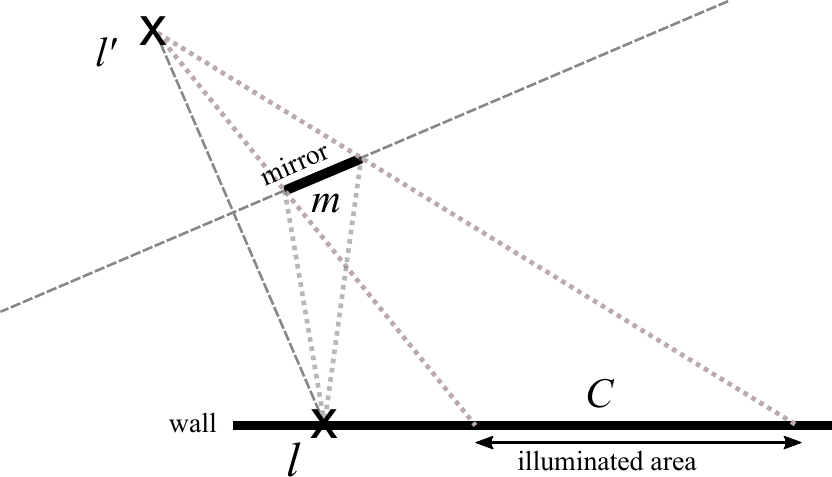}
\par\end{centering}
\caption{
To assess the optical path $\L \rightarrow \M^r \rightarrow \C$, we use a similarity relation: The laser spot $\L$ illuminates the wall as if it was reflected on the mirror plane, resulting in a virtual light spot $\L'$.
}
\label{fig:wall_illumination}
\end{figure}

For the time-of-flight computation we need to compute the length of a path $\S_L \rightarrow \L \rightarrow \M \rightarrow \C \rightarrow \S_C$. Note that $\M$ is a plane while $\S_L$, $\S_C$, $\L$, and $\C$ are points. Due to the specularity constraint of the mirror reflection there exists a unique point $\M^r$ on $\M$ at which the light is reflected. The length of the sub path $\L \rightarrow \M^r \rightarrow \C$ is equal to the path length $\L' \rightarrow \C$, where $\L'$ is the point $\L$ mirrored at $\M$ (see Figure~\ref{fig:wall_illumination}).

A mirror plane is represented in the Hesse normal form as normal vector $n$ and scalar offset $d$. Then:

\begin{equation}
\L' = \L - 2(n \cdot \L + d)n
\end{equation}

The total path length is the sum of all path segments:

\begin{equation}
f\left(S, \L, \C, \M\right)=\left\Vert \L-S_L\right\Vert +\left\Vert \C-\L'\right\Vert +\left\Vert S_C-\C\right\Vert 
\end{equation}

While mathematically planes are infinite, real mirrors are usually not. If $\M^r$ does not lie on the physical mirror plane, $\C$ will not receive any signal (see Figure~\ref{fig:wall_illumination}). In this case, the path can simply be removed from the optimization (see section~\ref{sec:calibration}).

\subsection{Calibration}
\label{sec:calibration}

Following the analysis-by-synthesis scheme, we take time-of-flight measurements $t$ from the real setup and compare them to the times-of-flight computed from the current estimate of the setup. If all possible combinations are used there are a total of $\#\Lset \cdot \#\Mset \cdot \#\Cset$ measurements. The target function of the optimization is the $\textrm{L}^2$ norm of the difference between the measurements:

\begin{equation}
\label{eq:opt}
\underset{S, \Lset, \Cset, \Mset}{\arg\min}\sum_{\L\in \Lset}\sum_{\C \in \Cset}\sum_{\M \in \Mset} \left\Vert f\left(S,\L,\C,\M \right)-t_{\L,\C,\M} \right \Vert ^{2}
\end{equation}

The calibration problem now consists of determining all $\L \in \Lset$, $\C \in \Cset$, $\M \in \Mset$, as well as $\S$, such that the corresponding optical path length matches the measured time of flight.

A calibration is only unique up to a rigid transformation of the whole setup as by definition a rigid transformation does not change any path lengths. A reconstruction obtained from such a transformed setup will differ only by said transformation (except for slight differences that can result from things like an axis-aligned reconstruction volume). We can therefore w.l.o.g define the camera and laser location $\S$ as the origin of the coordinate system and determine all other points relative to it. In general we consider the offset between the camera location $\S_C$ and the laser location $\S_L$ as a known feature of the hardware setup\footnote{Relative to the distance to the wall, the offset between $\S_C$ and $\S_L$ is usually small. In these cases the angle between $\S_C$ and $\S_L$ viewed from any $\C$ or $\L$ is marginal and the dominant factor is the total distance from the hardware to the wall.}.

Equation~\ref{eq:opt} assumes a \emph{fully connected} case where every possible path is used. Alternatively some paths can be omitted e.g. for confocal setups or when not all mirrors are used will all laser positions. The fully connected case is the default.

We solve Equation~\ref{eq:opt} with a standard gradient descent algorithm (BFGS~\cite{Tang:2014}). 
The initialization is further discussed in Section~\ref{sec:synthetic_evaluation}. Due to the compact image formation model automatic differentiation can be used for gradient computation.

\subsection{Parameterization}
\label{sec:parameterization}

In the most general case, each $\L$ and $\C$ has three components ($x,y,z$), while each $\M$ has four (represented in the Hesse normal form). This parameterization of $\M$ as a 4D vector does not guarantee unit length of the plane normal, so $f$ is evaluated on re-normalized planes.

From this general case more specialized parameterizations can be derived by applying a parameterization function $g: p \rightarrow (\S, \Lset, \Cset, \Mset)$ before $f$ is evaluated. We implement two of such parameterizations for common special cases. A suitable parameterization can decrease the degrees of freedom of the optimization (making it faster and more robust) and enforce certain constraints on the solution. Thus the usage of specialized parameterizations is in general preferred.

\subsubsection{Planar Walls}

To the best of our knowledge, all non-line-of-sight reconstruction approaches have used purely planar walls. Even though our method can be used for more general cases with the default parameterization, planar walls are thus an important special case.

After defining two basis vectors and an origin, each point on a planar wall can be described by just two components $u$ and $v$. As a calibration is only unique up to a rigid transformation we can w.l.o.g define the wall plane as the $X/Z$ plane. Then the only remaining parameter of the wall plane is the offset to our origin $\S$. As the mirrors reside outside the plane, their parameterization remains unchanged.

An initialization of the general case can be automatically transformed to a planar wall initialization by fitting a plane through $\Cset$ and $\Lset$ and transforming it onto the default plane (while also estimating an initial plane offset).

Due to its ubiquity and relevance, the planar wall parameterization is the default case in our evaluation.
 
\subsubsection{Regular Grids}

On two-dimensional camera sensors the individual pixels are usually arranged on a regular grid. This grid is projected into the scene along the view direction leading to strong constraints between the relative positions of the projected pixels. In the case of a planar wall this projection can be fully characterized by a homography, a $3 \times 3$ matrix (with a constant 1 as lower right entry) that maps homogeneous 2D coordinates of the image sensor to 2D coordinates on the wall.

Via a sensor pattern (a list of pixels in sensor coordinates) for projection, arbitrary layouts can be used. These are characterized by horizontal and vertical resolution of the sensor, 
non-square pixels, masking of dead pixels, lens distortion, and other potential parameters, that are hardware specific. We consider such characteristics to be given.

Since 2D sensors usually contain hundreds or thousands of pixels, the reduction of degrees of freedom  to a constant of 9 (8 for the homography, 1 for the wall plane) is significant. 
\section{Method evaluation}
\label{sec:synthetic_evaluation}
For practical usage of the calibration it is important to understand on which setups it can be applied and what the expected quality of the result is. Since the optimization problem can be formulated for every setup, we define robustness of the solution as criterion for applicability.

A setup is characterized by a number of parameters, some of which are easier to change than others. Fixed parameters include those defined by the hardware, e.g., the resolution of the image sensor (the number of camera pixels) and the accuracy of the time-of-flight information. 
Flexible parameters include the number of laser positions, the number of mirror positions and the quality of the initialization. 
Note, that even if the setup will later only use a single laser position, it is typically possible to use additional laser positions for calibration measurements.

In the evaluation we give guidelines on how these flexible parameters should be chosen before we show the expected accuracy depending on the fixed parameters.
\subsection{Evaluation setup}
\label{sec:evaluation_setup}
\begin{figure}
\begin{centering}
\includegraphics[width=0.7\columnwidth]{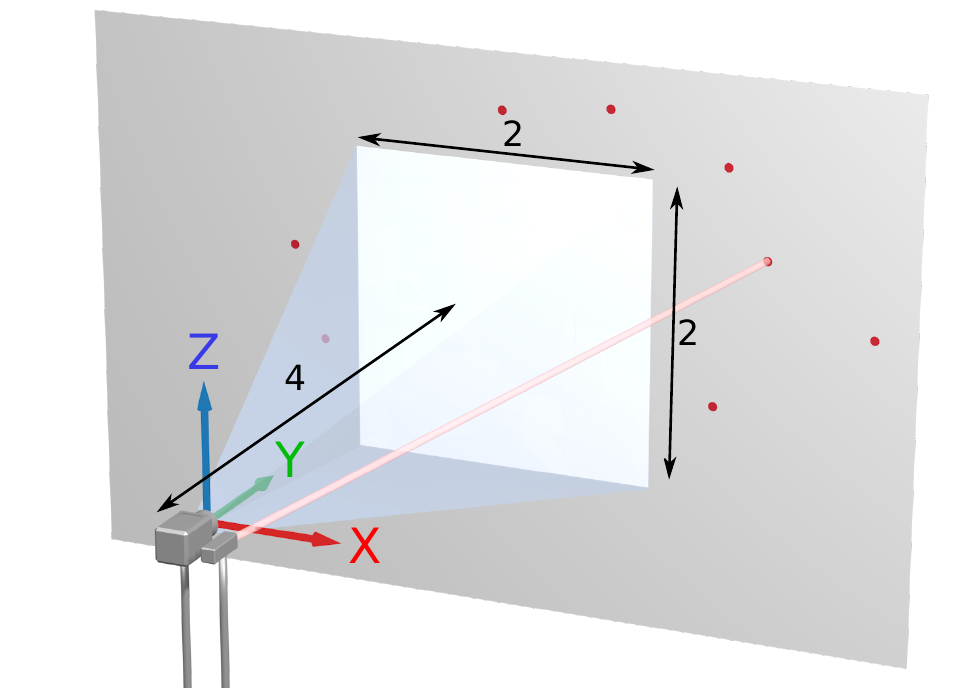}
\par\end{centering}
\caption{\label{fig:synthetic_setup}Setup used for the synthetic evaluation. The camera and laser are in the origin, the red dots mark the laser spot positions. The in total 40 mirrors (not shown here) are placed between camera and wall and face the later.}
\end{figure}
Our standard evaluation setup consists of 25 camera pixels (arranged in a $5 \times 5$ grid), 8 laser spot positions and 40 mirror positions. During the evaluation a varying amount of the laser and mirror positions are used. The camera view frustum on the wall is $2\times2$ units and 4 units away from the camera and laser. The laser spot positions are arranged around the view frustum while the mirrors are placed in front of the wall. 
We use the default case of a planar wall for the majority of the evaluation. The setup is shown in Figure~\ref{fig:synthetic_setup}.
To mimic real calibration situations, we rely on the ground truth setup and apply varying levels of noise to resemble measurement uncertainties. 
This perturbed data is then used as initialization for the optimization process, which helps us to assess what level of accuracy is required to successfully estimate the correct geometry.
In particular, we apply measurement noise with standard deviation $n$ as

\begin{itemize}
	\item Gaussian noise with standard deviation of $n$ to pixel and laser spot positions,
	\item Gaussian noise with standard deviation of $n/4$ to mirror normals and renormalize them,
	\item and Gaussian noise with standard deviation of $n$ to the mirror plane offsets.
\end{itemize}

Figure~\ref{fig:teaser} shows the ground truth values along with an example initialization where spatial noise with a standard deviation of $n=0.5$ was applied. At this noise level not much of the original structure is preserved. Equally, Gaussian noise in various levels is applied to the reference time-of-flight values $t$.

We characterize the quality of a calibration by the root-mean-square (RMS) error between the individual components. Mirror positions are not considered part of the calibration result and thus excluded from evaluation.
For two setups $P=\{\S_1, \L \in \Lset_1, \C \in \Cset_1\}$ and $Q=\{\S_2, \L \in \Lset_2, \C \in \Cset_2\}$ we compute:
\begin{equation}
\label{eq:setup_rms}
\textrm{RMS}(P, Q) = \sqrt{\sum \left\Vert P_i - Q_i \right\Vert_2^2}
\end{equation}
As the calibrated setup might be in a different coordinate system, naively applying Equation~\ref{eq:setup_rms} can result in high errors even for actually good result. 
Therefore we determine an optimal rigid transformation that transforms a setup onto a reference, after which the RMS becomes meaningful. 
Since the time-of-flight values for individual paths are in a defined order, correspondences between setup points remain intact during optimization and the optimal rigid setup transformation can be computed with the Kabsch algorithm~\cite{Kabsch:1976}. Since the RMS error has the same unit as the initialization noise $n$, the two can directly be set into relation.

The example in Figure~\ref{fig:teaser} uses 4 mirror positions and  and time-of-flight noise with a standard deviation of 0.02 was applied. It achieves a reconstruction error of 0.042.

\subsection{Required Measurements}
\label{sec:required_measurements}
\begin{figure}
\begin{centering}
\includegraphics[width=1\columnwidth]{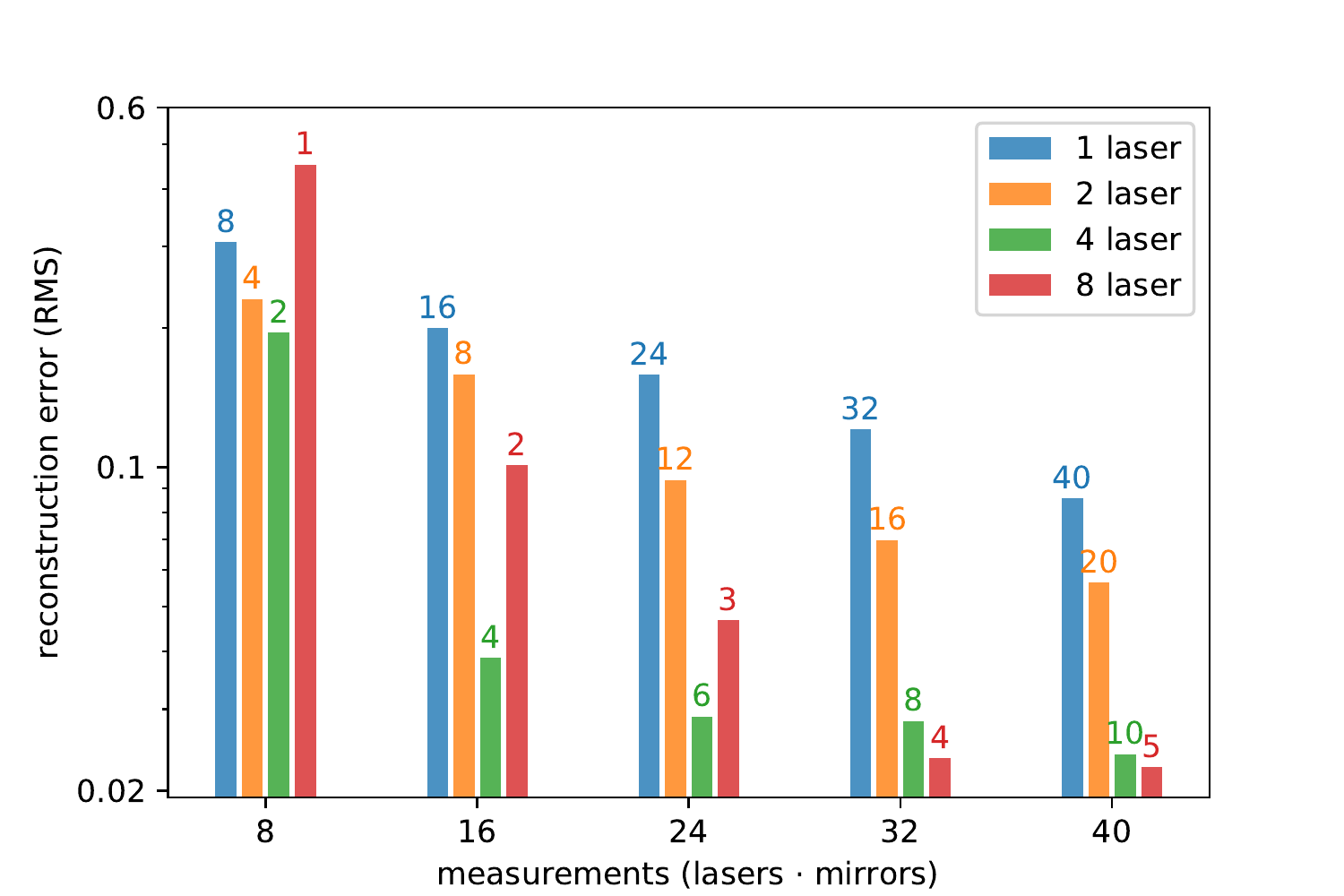}
\par\end{centering}
\caption{\label{fig:measurements}Calibration performance depending on the total number of measurements. Each data point shows the mean and standard deviation of 100 individual optimizations. The numbers above the bars show the number of mirrors used for that data point.}
\end{figure}

For a robust optimization the ratio between the input dimensions and output dimensions is an important measure. 
The number of input dimensions of the optimization problem is defined by the amount of measurements (i.e., used paths), 
while the number of output dimensions depends on the parameterization. 
For the fully connected case there are $\#\Lset \cdot \#\Mset \cdot \#\Cset$ measurements: All possible connections between laser, mirror and camera are included.
The output dimensions are (when the origin is set to $\S$):
\begin{itemize}
	\item Default: $3\cdot \#\Cset + 3 \cdot \#\Lset + 4 \cdot \#\Mset$
	\item Planar: $2\cdot \#\Cset + 2 \cdot \#\Lset + 4 \cdot \#\Mset + 1$
	\item Grid: $2 \cdot \#\Lset + 4 \cdot \#\Mset + 9$
\end{itemize}
In Figure~\ref{fig:measurements} we compare the total number of measurements to the reconstruction error, where the RMS error is color-coded with respect to the number of lasers involved in the measurement. 
To analyze the performance for different combinations of laser and mirror positions, we split the same number of measurements across lasers and mirrors to different parts. 
All optimizations use the planar parameterization and are initialized with a noise level $n \in [0 \ldots 0.5]$ and a time-of-flight noise level of $0.02$.

The results show that the number of measurements alone says little about the structure of the problem, the same number of measurements may lead to severely different errors depending on the ratio between lasers and mirrors used. 
As expected, the reconstruction improves when more measurements are used. More interestingly, it is also beneficial to have about as many laser positions as there are mirror positions: The more extreme the ratio between laser and mirror positions, 
the worse the results become.
For practical applications, the reconstruction error should be close to or below the depth resolution of the camera. We conclude that 32 measurements using at least 4 laser positions are a lower bound for a sufficiently accurate reconstruction.

\begin{figure}
\begin{centering}
\includegraphics[width=1\columnwidth]{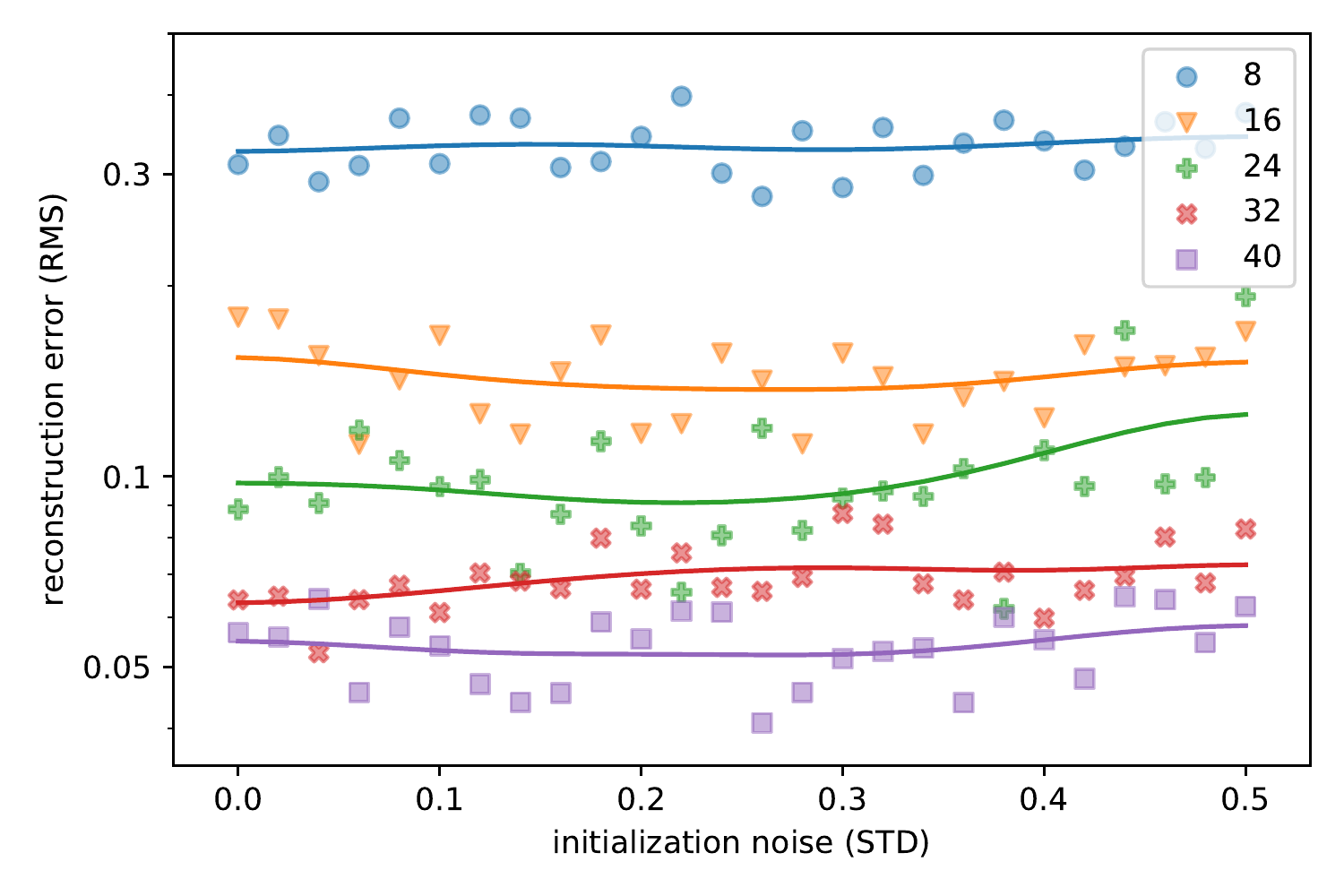}
\par\end{centering}
\caption{\label{fig:initialization}Reconstruction error for various levels of initialization noise. The data is the same as shown in Figure~\ref{fig:measurements}, averaged over all laser/mirror combinations for a specific number of measurements (shown in different colors / markers).}
\end{figure}

\begin{figure}
\begin{centering}
\includegraphics[width=1\columnwidth]{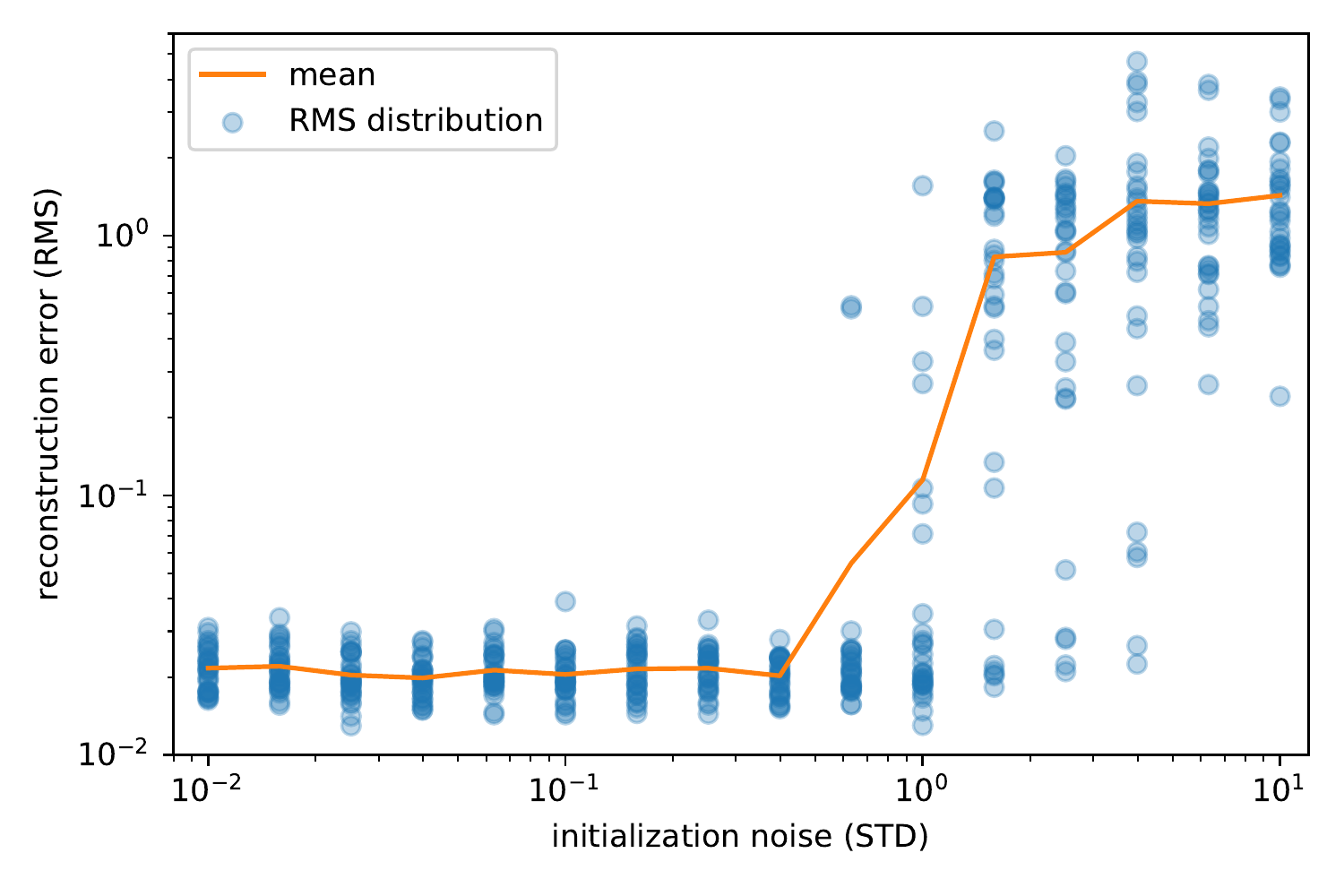}
\par\end{centering}
\caption{\label{fig:divergence}Reconstruction success depending on initialization noise. Time-of-flight noise is fixed at 0.02. The blue distribution of the individual optimization results gives an better intuition than the orange mean value - the results split in two distinct clusters for increased noise. Note that both axes are in log scale.}
\end{figure}

Figure \ref{fig:initialization} shows the same data set as in Figure~\ref{fig:measurements}, but this time decoded in terms of its dependence on the initialization error.
We find that within generous bounds the initialization has no effect on the convergence of the optimization; the RMS error primarily depends on the number of measurements involved.
Figure~\ref{fig:divergence} shows the limits of the allowed initialization error. Even for high values \emph{some} optimization runs still converge to the correct result, but there are no guarantees and it cannot be considered a safe initialization.

Up to a certain threshold close to $10^0$ units, the distribution of reconstruction errors is strongly centered at low RMSE, indicating an accurate result (note the log-log scale). 
Once this threshold is crossed, the optimization does not converge anymore and exhibits a sudden drop in quality. 

We can transfer these insights to form an important rule with respect to the calibration of real setups: We cannot rely on arbitrary initialization values but indeed require a rough knowledge of the geometry. Still, 
even a rough estimate is sufficient to yield a very accurate calibration, which might not even require the use of measuring tapes and rulers.

\begin{figure}
\begin{centering}
\includegraphics[width=1\columnwidth]{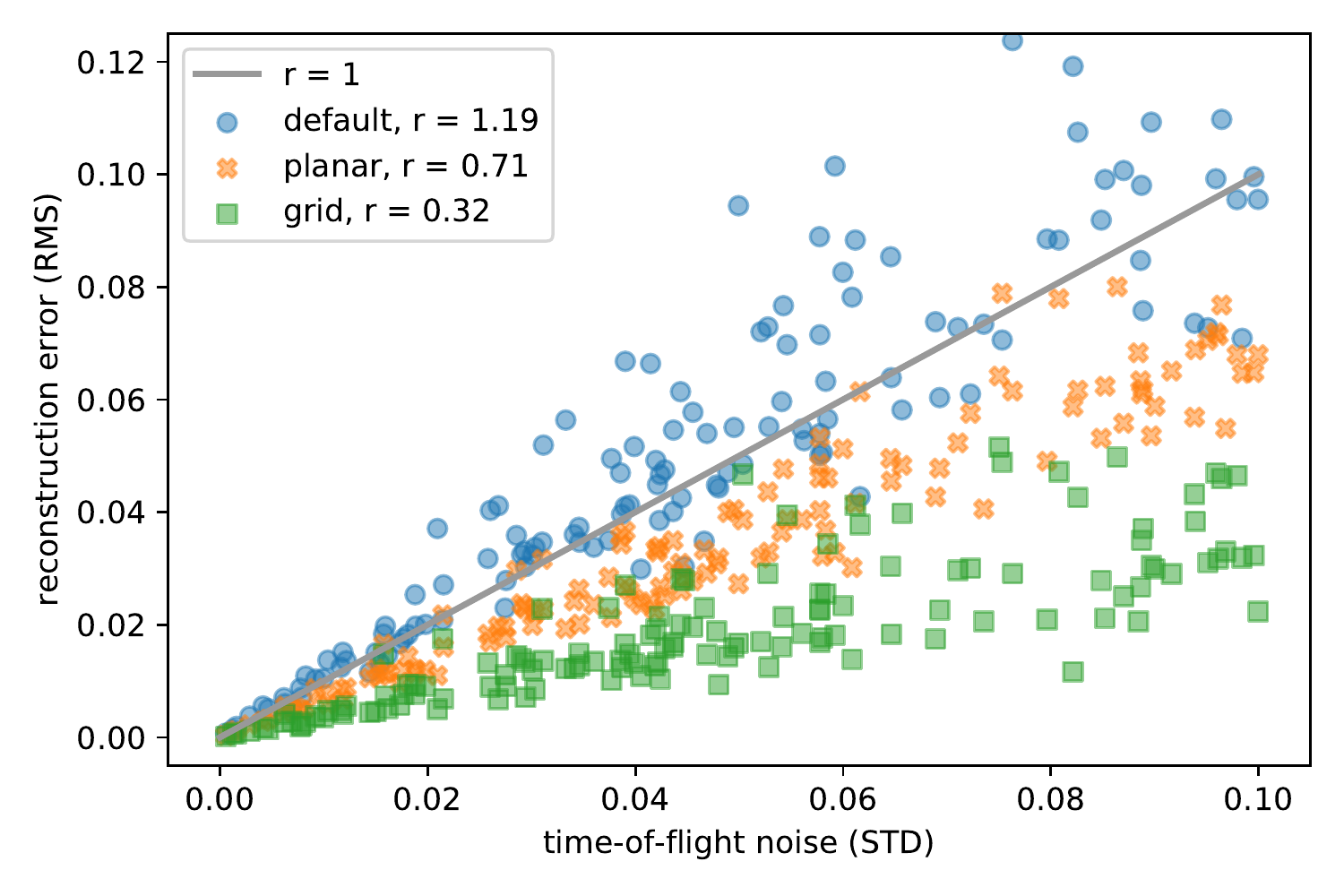}
\par\end{centering}
\caption{\label{fig:tof_sensitivity}Calibration error depending on the time-of-flight noise and parameterization. The r-values show the gradient of a linear fit for each parameterization.}
\end{figure}

Figure \ref{fig:tof_sensitivity} shows the reconstruction error in dependency of the time-of-flight noise and the parametrization. To generate the data the standard setup with 5 mirrors and 6 laser positions is initialized with a random noise value between $0$ and $0.5$. We find that there is an approximately linear relationship between the uncertainty of the time-of-flight data and the reconstruction error.

\begin{figure}
\begin{centering}
\includegraphics[width=0.48\columnwidth]{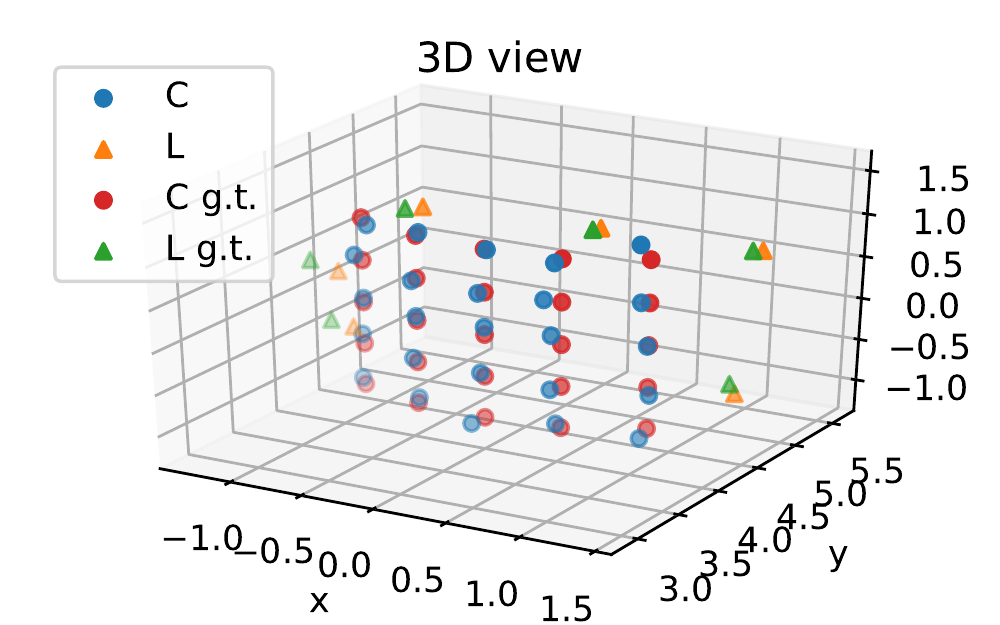}
\includegraphics[width=0.48\columnwidth]{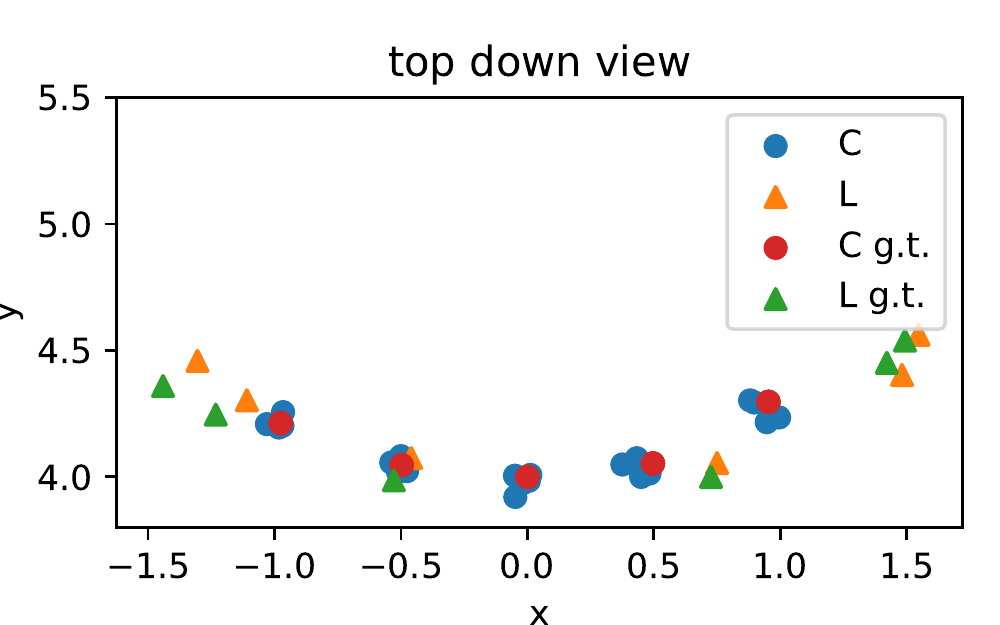}
\par\end{centering}
\caption{\label{fig:curved_wall}Example calibration of a curved wall. The setup consists of 6 lasers and 6 mirrors, the initialization noise is 0.5, the time-of-flight noise is 0.1. The RMS error of the calibration is 0.099.}
\end{figure}

The naive parameterization support arbitrary three-dimensional positions. An example calibration is shown in Figure~\ref{fig:curved_wall}.

Our main findings of this analysis are, that for a sufficient amount of measurements, a wide area of safe initialization exists. However the particular setup geometry also plays a certain role, e.g. if all laser positions are confined in a small part of the scene, the data is mostly redundant and the optimization is less stable. For optimal results, an equal amount of laser points and mirrors should be used, equally (but not symmetrically) distributed in the scene.

\subsection{Implementation and Runtime}
In our prototype Equation~\ref{eq:opt} is implemented purely in Python, the optimization is performed with the \texttt{scipy.optimize} package. On typical setups, the optimization runs for about 2 minutes, but obviously much more efficient implementations could be used if calibration speed becomes a concern.
 
\section{Experimental Results}

For the experimental verification of our results we calibrate a setup consisting of a PrincetonLightwave InGaAs Geiger-mode avalanche photodiode camera and a Keopsys pulsed Er-doped fiber laser. The camera has a spatial resolution of $32 \times 32$ pixel and a bin width of 250 ps (7.495~cm at $c$). Each measurement consists of 200,000 individual binary frames captured in about 4 seconds.
The laser emits light at a wavelength of 1.55~$\mu$m and has a pulse length of 500~ps. The transient histograms retrieved from the camera are converted to time-of-flight values by fitting a Gaussian function to the main peak.

\subsection{Calibration Results}
\label{sec:calibration_results}

Our setup uses a planar wall at a distance of 6.6\,m from the camera. The field of view on the wall is approximately 135\,cm~$\times$~135\,cm, resulting in a size of a projected pixel of 4.2~cm~$\times$~4.2~cm. We measured the cameras field of view using a moving marker on the wall and observing it in the cameras live image. The 7 spot positions of the near-infrared laser where measured using an IR detector card. We estimate that these measurements are accurate up to 1--2\,cm. The signal offset between camera and laser (which results in a time-of-flight offset) is calibrated by placing a planar calibration target in front of the setup at several known distances.

A household-grade mirror is mounted on a tripod which we place at 7 different locations in the scene. The mirror planes were initialized by measuring the position of the tripod over the floor and assuming that the plane normal faces towards the geometric mean of the camera and laser points. Although being a rough estimate, this approach proved sufficient.

Measurements are affected by scattering (e.g. when the laser spot is close to the view frustum or the laser beam crosses it and hits tiny particles in the air) and a couple of rows and columns on the image sensor are broken, resulting in invalid values. As our proposed method uses a flexible list of $\L \rightarrow \M \rightarrow \C$ paths, we can automatically detect and remove invalid paths from the optimization (see the supplementary material for details on the detection). Due to the high amount of pixels and their regular layout, we use the grid parameterization.

\begin{figure}
\begin{centering}
\includegraphics[width=0.48\columnwidth,align=c]{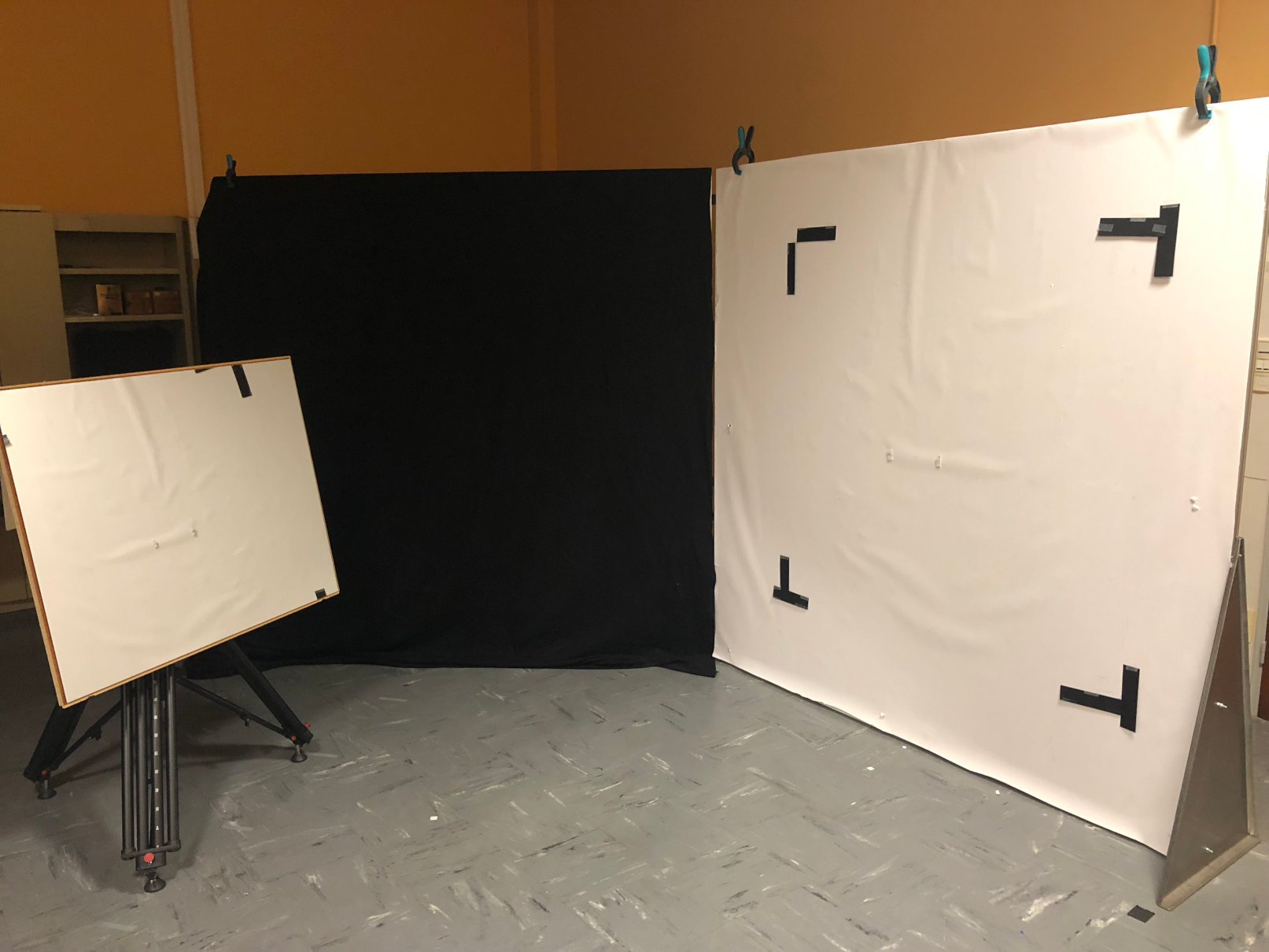}\hfill
\includegraphics[width=0.48\columnwidth,align=c]{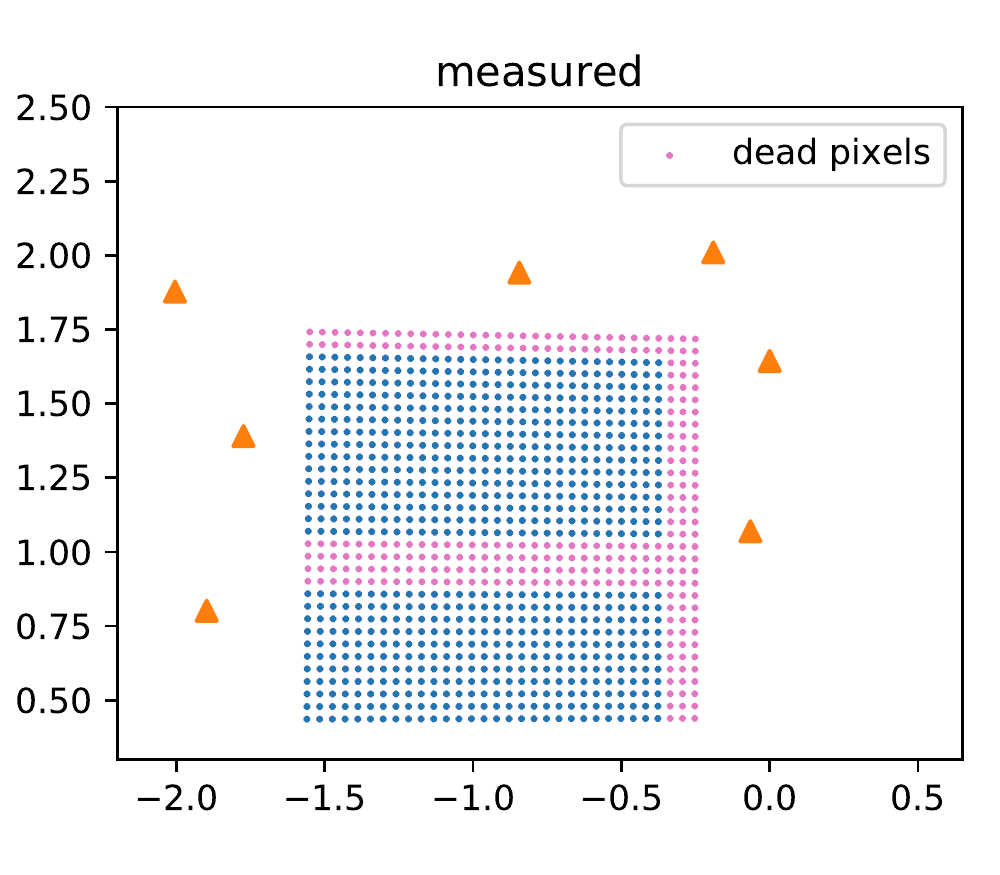}

\includegraphics[width=0.48\columnwidth,align=c]{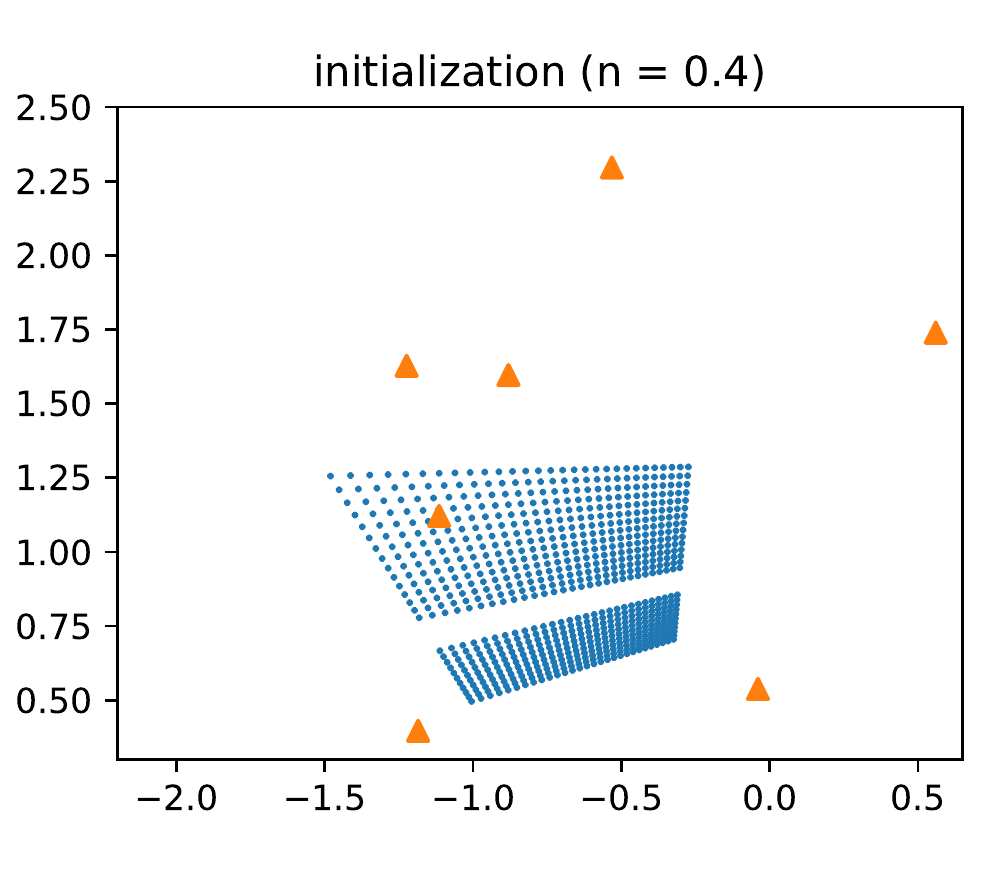}\hfill
\includegraphics[width=0.48\columnwidth,align=c]{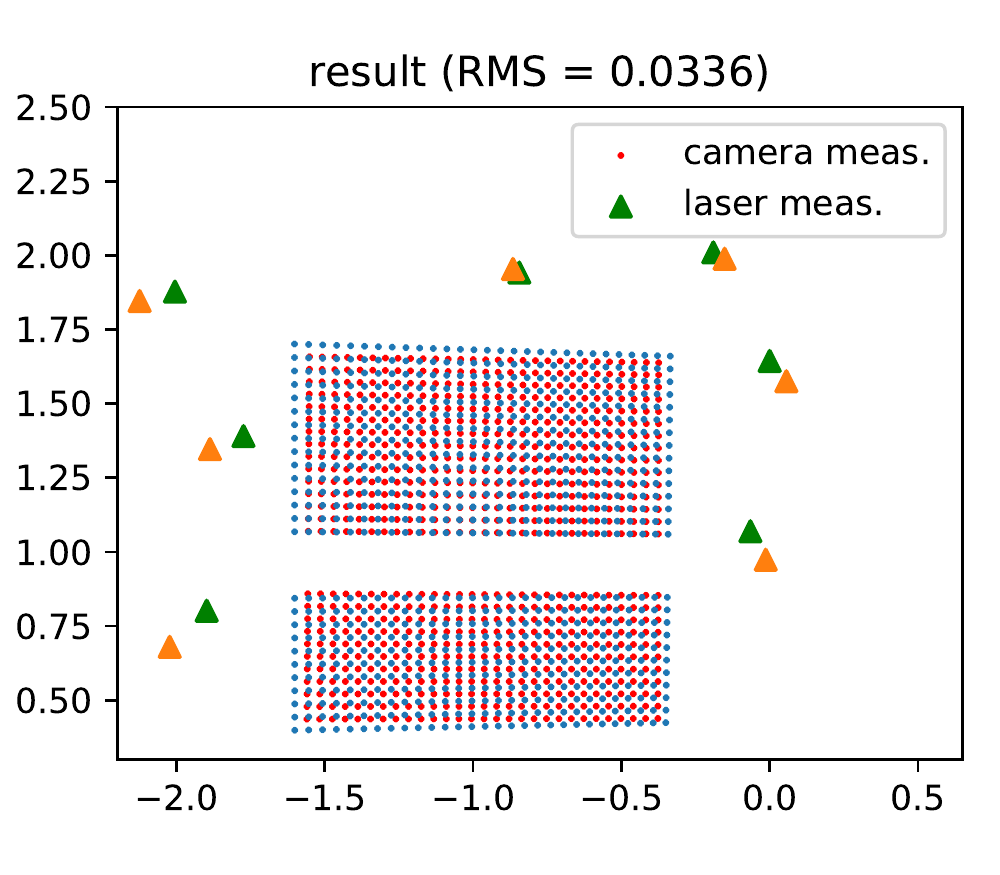}
\par\end{centering}
\caption{\label{fig:experimental_result}Calibration result on measured data. The calibration target mirror is mounted on a tripod in front of the reflector wall (top left). Some broken camera pixels were removed (top right). The initialization noise of 40\,cm (bottom left) is almost a third of the camera field of view. The RMS of the reconstruction (bottom right) is 3.4~cm.}
\end{figure}

Figure~\ref{fig:experimental_result} shows a typical calibration result. Our measurement of the setup geometry acts as estimated ground truth comparison and the optimization was initialized as described in section \ref{sec:evaluation_setup} with an average error of $n=$~40\,cm. Note, that since the grid parameterization is used, noise is applied to the corners of the view frustum instead of individual pixels (since their layout is given by the sensor pattern). In real applications such an rough initialization could be estimated without the use of measuring tape or similar devices.

Over multiple optimizations we achieve a typical RMS error of 3--4\,cm on this setup. Considering the bad temporal resolution of the setup, these results are consistent with our findings insection~\ref{sec:synthetic_evaluation}. With twice the temporal resolution (which is readily achievable with commonly used hardware) we would expect a reconstruction error on the order of the uncertainty of our careful manual measurements.

\subsection{Backprojection Results}

\begin{figure}
\begin{centering}
\includegraphics[width=0.48\columnwidth,align=c]{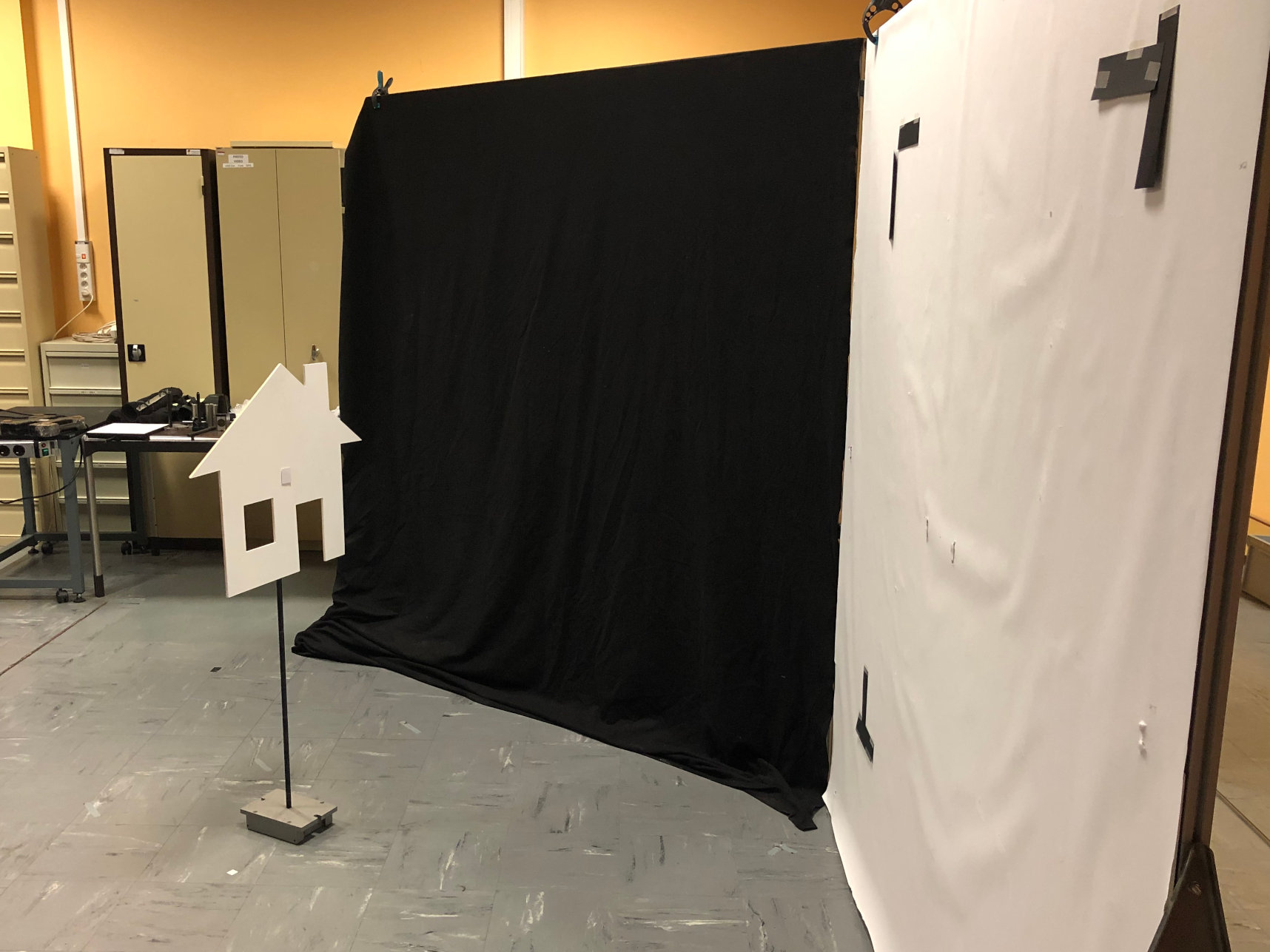}\hfill
\includegraphics[width=0.48\columnwidth,align=c]{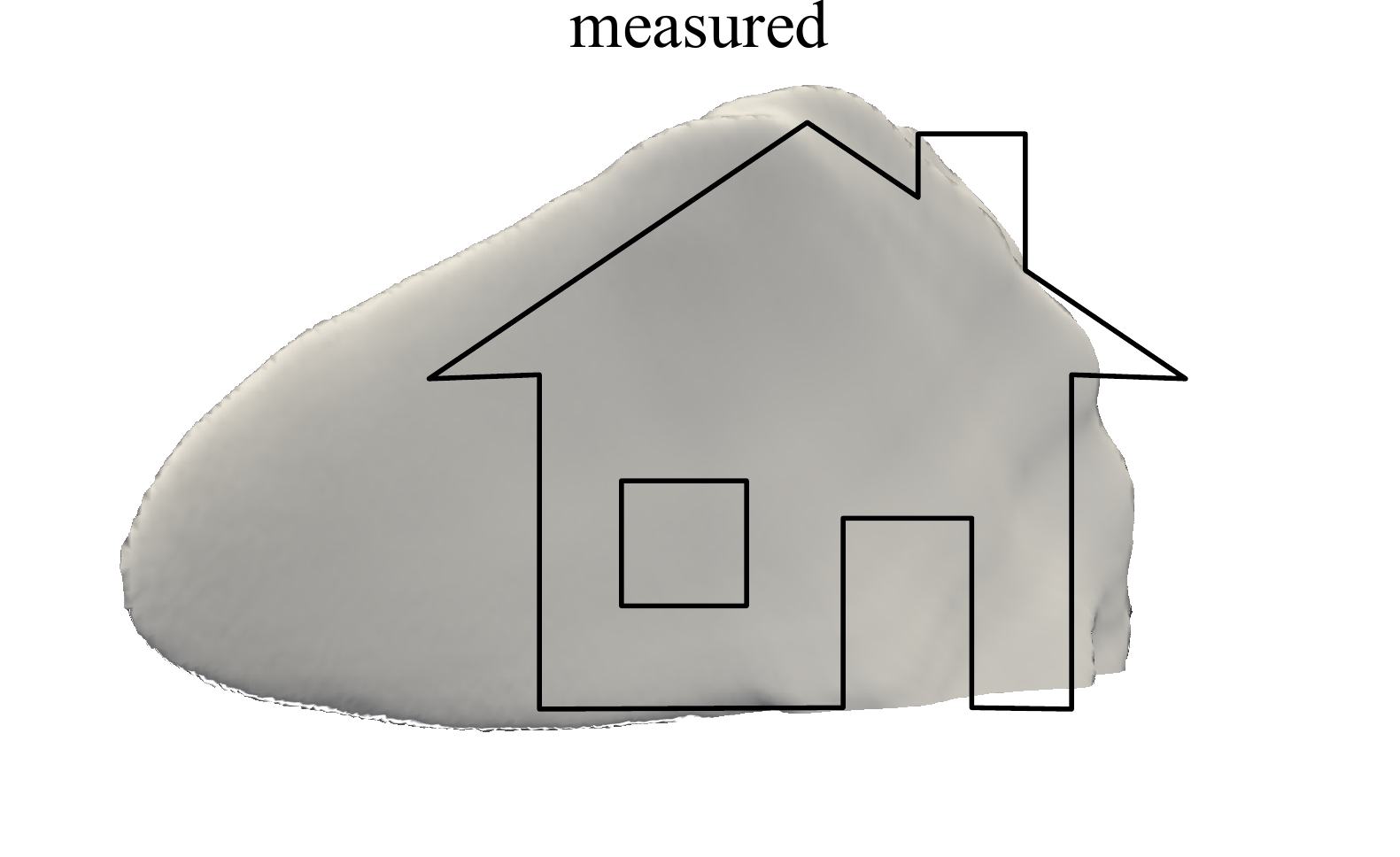}
\vspace{1em}

\includegraphics[width=0.48\columnwidth,align=c]{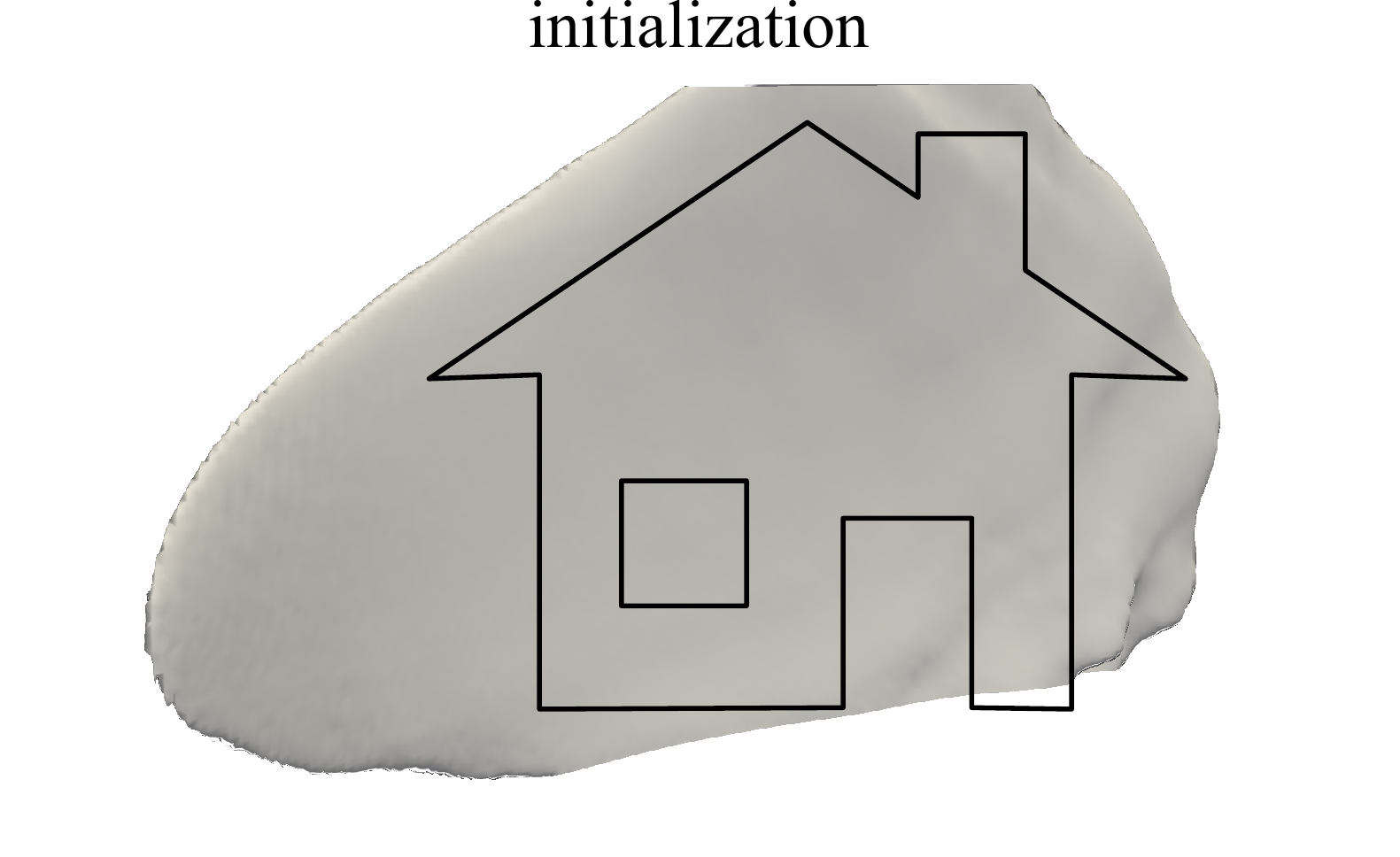}\hfill
\includegraphics[width=0.48\columnwidth,align=c]{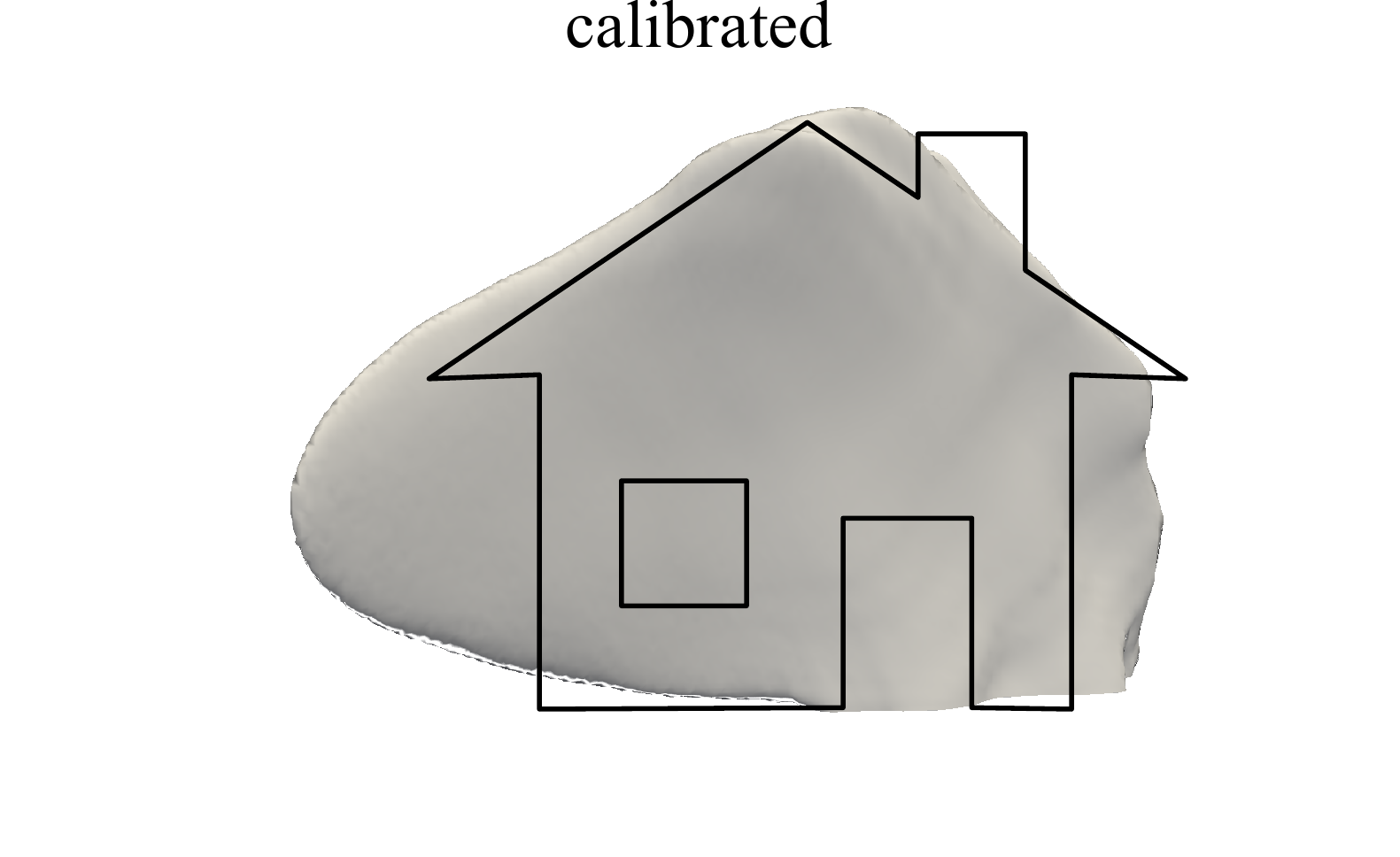}
\par\end{centering}
\caption{\label{fig:experimental_backprojection}Visual comparison of the backprojection results. The calibration is initialized with a standard deviation of $n=10$ cm (with an object size of 56.9 cm $\times$ 44.1 cm).}
\end{figure}

We test the performance of the calibration by comparing reconstruction results from uncalibrated setups with results from the calibrated ones. A house-shaped wooden target is placed in front of the wall (see Figure~\ref{fig:experimental_backprojection}) and reconstructed using a backprojection implementation as described in Liu et al.~\cite{liu2019non}. Note that the depth resolution of our setup is significantly lower than in Liu et al.~\cite{liu2019non} and that we did not fine-tune parameters for individual reconstructions in order to show the differences caused by the different setup geometries. The same setup as in Section~\ref{sec:calibration_results} is used.

While none of the reconstructions in Figure~\ref{fig:experimental_backprojection} reveal fine details of the object, there are significant differences between the reconstructions. After noise is applied to the measured ground truth (as described in section \ref{sec:evaluation_setup}) the ellipsoids in the backprojection are less focused and result in a blurred-out shape. The calibrated setup has a RMS distance of 3.1\,cm to the measured setup, the precise ground truth value is unknown but in the vicinity of both. The reconstruction from the calibrated setup appears more focused than the other two.

\section{Conclusion}

Our proposed method for non-line-of-sight setup calibration is demonstrated to robustly optimize real-world setups. Despite being a non-convex problem we show that a generous convergence basin exists around the global minimum which results in low requirements of the initialization. The achieved accuracy depends on the depth resolution of the setup, but setup specific parameterizations can be used to enforce constraints and increase the accuracy. As the mirror target results in a single sharp peak in the signal, we do not rely on hardware being able to record full transient histograms. This makes our method applicable on a wide variety of hardware including amplitude-modulated continuous-wave lidars. The ability to calibrate also non-planar walls could enable non-line-of-sight imaging applications in everyday situations.

Future work includes evaluation and fine tuning of the calibration for a broader set of setups such as confocal ones. Additionally the mirror that acts as calibration target could be augmented with a calibration pattern that is then projected onto the wall. This would allow to capture additional information which could possibly be used to improve results. Similarly using also the intensity of paths could allow to formulate additional constraints on the wall normal. 
\section{Acknowledgments}
This work was supported by the German Research Foundation (DFG) under
grant HU-2273/2-1 and the ERC starting grant “ECHO”. We thank Sebastian Werner for helpful discussions and feedback.

{\small
\bibliographystyle{ieee_fullname}

}

\end{document}